\documentclass[twocolumn,aps,prb,superscriptaddress]{revtex4-2}

\usepackage{graphicx}
\usepackage{amssymb}

\begin{document}

\preprint{preprint}

\title{Evidences for magnetic dimers and skyrmion lattice formation in Eu$_2$Pd$_2$Sn}


\author{J.G. Sereni}
\email[]{jsereni@yahoo.com}
\affiliation{Low Temperature Division, CAB-CNEA, CONICET, IB-UNCuyo, 8400 Bariloche, Argentina}

\author{I. Čurlik}
\affiliation{Faculty of Sciences, University of Prešov, 17. novembra 1, SK - 080 78 Prešov, Slovakia}

\author{M. Reiffers}
\affiliation{Faculty of Sciences, University of Prešov, 17. novembra 1, SK - 080 78 Prešov, Slovakia}
\affiliation{Institute of Experimental Physics, Slovak Academy of Science, Watsonova 47, Košice, Slovakia}

\author{M. Giovannini}
\affiliation{Department of Chemistry, University of Genova, Via Dodecaneso 31, Genova, Italy}

\date{\today}

\begin{abstract}
Magnetic, thermal and transport properties of the non-centrosymmetric compound Eu$_2$Pd$_2$Sn are 
revisited after including  new measurements. In its paramagnetic phase, the outstanding feature of this compound is the formation of Eu$^{2+}$ dimers that allows to understand the 
deviation of the 
magnetic susceptibility $\chi(T)$ from the C-W law below about 70\,K, the field dependent magnetization 
$M(B)$ variation below $\approx 80$\,K and the reduced entropy at the ordering temperature $S(T_N)= 0.64R\ln(8)$. A significant change of the exchange interactions occurs between 
$T\approx 70$\,K (where 
$\theta_P = 18$\,K) and $T_N = 13.3$\,K (where $\theta_P =-4.5$\,K). 
The strong electronic overlap, arising from the reduced Eu-Eu spacing compared with that of pure Eu$^{2+}$ is expected to power these quasiparticles formation, inducing a 
significant reformulation of the magnetic structure. 

A rich magnetic phase diagram is obtained from the analysis of the derivatives of the magnetic parameters: $\partial \chi/\partial T$ and $\partial M/\partial B$, and the field dependence of the 
specific heat. 
Two critical points are recognized and a tentative description of the magnetic structures is proposed.
The possible formation of a skyrmion lattice, that arises from the presence of magnetically frustrated pockets in the phase diagram, is suggested by theoretical studies on hexagonal structures exhibiting  
similar interactions pattern.

\end{abstract}


\maketitle


\section{Introduction}

Ternary magnetic rare earths (RE) compounds mostly exhibit ground states (GS) with low degeneracy owing to the effect of crystalline electric field (CEF). Depending on the integer or half-integer character of its total angular momentum $J$ and the local symmetry, their GS do not exceeds a fourfold degeneracy (e.g. in cubic structures). Consequently, their effective magnetic moments $\mu_{eff}$ at low 
temperatures are clearly reduced in comparison to 
those at high temperatures, unless a magnetic order occurs at higher temperature than the splitting induced by the CEF. Nevertheless, such possibility is quite unlikely in ternary compounds because of the typically large spacing between magnetic atoms. 

The well known exception is the pure spin Gd$^{3+}$ with [Xe][6s$^2$5d$^1$4f$^7$] configuration because, with its $J = S = 7/2$ and orbital momentum$L = 0$ , it is not affected 
by the symmetry reduction produced by the CEF. There is another exception: Eu$^{2+}$ in its excited electronic configuration [Xe][6s$^2$4f$^7$] that, for magnetic purposes, it is usually 
considered equivalent to the Gd$^{3+}$. However, the configuration of Eu$^{2+}$  
is not identical to Gd$^{3+}$ despite to have the same number of $4f$ electrons.

In fact, such difference is reflected in some physico-chemical properties like the respective atomic radii: 2.04\,$\AA$ for Eu$^{2+}$ and 1.80\,$\AA$ for Gd$^{3+}$, 
as well as their temperature and heat of melting \cite{melt}. 
As a consequence, Eu may form a number of compounds which are normally allowed for large divalent alkaline 
earths (e.g. Ca$^{2+}$), but not for smaller Gd$^{3+}$ atoms. Nonetheless, to our knowledge there are no studies comparing the actual Eu(Z=63) [Xe] [6s$^2$4f$^7$] electronic “form factor” with that  of Gd(Z = 64).

\begin{figure}
\begin{center}
\includegraphics[width=22pc]{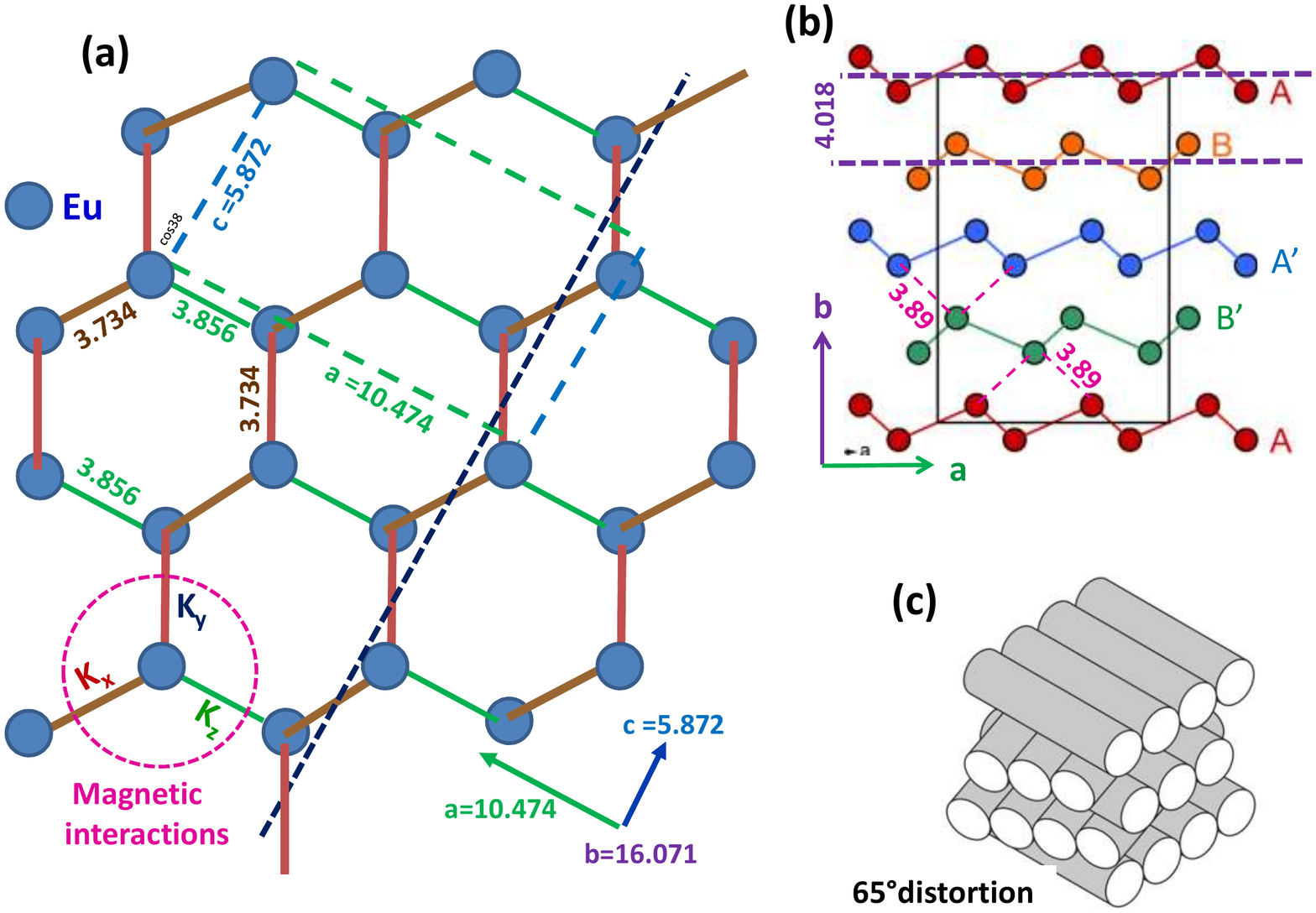}
\caption{(Color online) (a) View of the Eu atoms network projected on the 'ac' plane \cite{Doverbratt}, including Eu-NN (brown segments) and Eu-NNN (green segments) distances: 
3.734\,$\AA$ and 3.856\,$\AA$ respectively \cite{ours}. The blue-dashed long line connects the crossing points between Eu-NN lines with the 'ac' plane. The green-dashed rectangle outlines the unit cell,  and the magenta dashed circle contains the neighboring magnetic interactions labeled as $\sf K_{x,y,z}$ according to a Kitaev representation \cite{Kitaev}. 
(b) Zig-zag Eu chains staked along the 'b' direction and projected on the 'ab' plain (dashed brown lines).
(c) Cylinder-like structural fragments representing Eu chains wrapping respective linear Pd chains (after Fig.~\ref{F1}I-b of Ref.\cite{ours}) and forming a 65° dihedra between layers. 
\label{F1}}
\end{center}
\end{figure}

Besides that, the lack of CEF effect allows both RE to exhibit the largest $\mu_{eff}$ values. This is the reason why there is a growing interest in the search of new Eu$^{2+}$ ternary compounds \cite{Poettgen, Seiro, 1-Rh-3, RyanGe3, EuNi5As3, Diego, 1-1-2, ours, EuIr2P2}. Among them, we 
revisit in this work the already 
characterized Eu$_2$Pd$_2$Sn \cite{ours} in order to better understand its complex magnetic behavior and to built up a magnetic phase diagram. For such purpose new measurements were carried including detailed field dependent magnetic susceptibility, magnetization and low temperature magnetoresistence.

One of the remarkable features of the crystal structure of  Eu$_2$Pd$_2$Sn (Ca$_2$Pd$_2$Sn-type 
\cite{Doverbratt}) is the non-centro-symmetric character of its 
orthorhombic structure, where magnetic Eu nearest neighbors (Eu-NN at 3.734\,$\AA$) form zig-zag chains, 
represented by brown lines connecting Eu atoms in Fig.~\ref{F1}a. These chains 
lie on the 'ac' plane, with Eu atoms alternatively displaced up and down respect to the 
plane, see Fig.~\ref{F1}b. There, one can see how the Eu-chains are stacked along the 'b' direction, where two consecutive chains in the 'b' direction belong to an AB bilayer while the following pair belong to an A'B' bilayer.  
This distinction between AB and A'B' bilayers arises from the fact the each bilayer wraps a linear Pd chain \cite{Doverbratt} which points to two alternate directions, schematically depicted in Fig.~\ref{F1}c, with Eu and Sn atoms disposed in entangled chains represented by parallel cylinders. Each cylinder, with 
respective axes defined by Pd linear chains, contains upper and lower Eu zig-zag chains forming the 
bilayers indicated in  Fig.~\ref{F1}b (e.g. AB or A'B'), while consecutive cylinders forms a 65° dihedra \cite{ours} as shown in Fig.~\ref{F1}c. 

Coming back to Fig.~\ref{F1}a, one can see that all neighboring chains lying in the same ‘ac’-plane form a network of puckered-elongated hexagons. The elongated sides (green 
segments in the figure) correspond to the next nearest Eu-neighbors (NNN at 3.856\,$\AA$). Those puckered networks are stacked in the ABA’B’ layers sequence previously mentioned \cite{Doverbratt}.   
Notice that, despite the mean distance between layers (L) is $LL=4.018\AA$, see Fig.~\ref{F1}b, due to the 
puckered topology of the hexagonal network there are two Eu third neighbors at $Eu_{LL}=3.89\AA$, both 
alternatively belonging to the upper and lower layer. In other words, if Eu$_i$ atoms of chain B' in 'even' positions ($i=0,2,4, ....$) have both $Eu_{LL}$ atoms on the upper (A') layer,  
the 'odd' atoms ($i=1,3,5, ....$) have them on the lower (A) layer which belongs to the following bilayer.  

There is a further structural peculiarity in this compound. As mentioned before the Eu-NN spacing is $3.734\,\AA$ whereas the corresponding distance in pure Eu$^{2+}$ metal is $4.08\AA$ \cite{radius}. The strength of this electronic overlap can be evaluated by comparing this difference: $(4.08-3.734)/4.08 \approx 8.5\%$, with the one between Eu$^{2+}$ and Eu$^{3+}$ metals: $\approx 11\%$. Notably, despite such  electronic overlap Eu atoms keep 
their expanded Eu$^{2+}$ configuration which is recognized in its magnetic behavior. To our knowledge, this unique property is not found in any other Eu compound and has relevant consequences in its magnetic structure.   

In this work we will firstly analyze the consequences of the mentioned peculiar Eu network on the magnetic and thermal properties in the paramagnetic phase, following with the those of the ordered phase 
whose tentative magnetic phase diagram will be constructed. 

\section{Experimental results}
\subsection{Paramagnetic Phase}

\subsubsection{Low field Magnetic Susceptibility at intermediate temperature}

\begin{figure}
\begin{center}
\includegraphics[width=20pc]{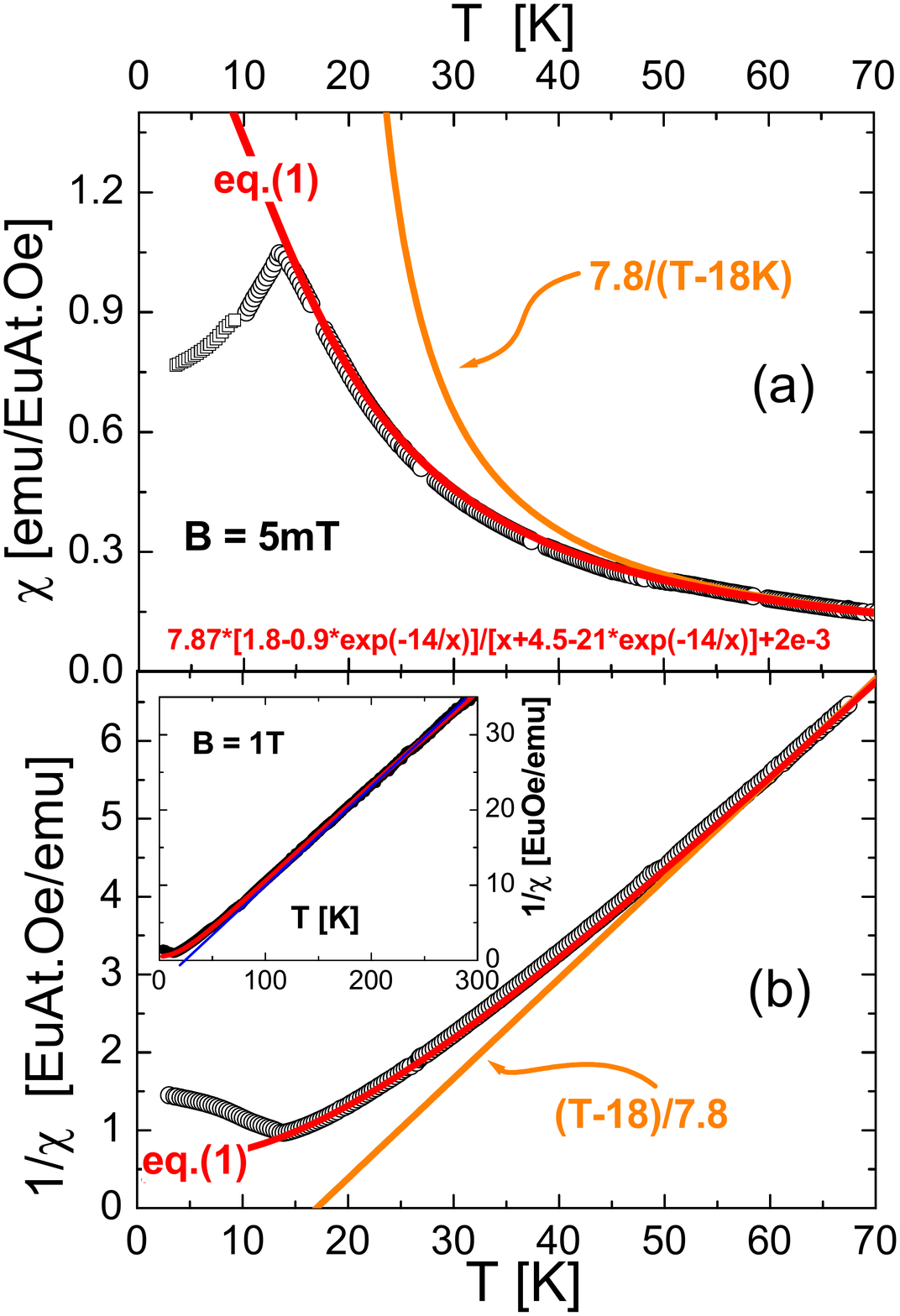}
\caption{(Color online) a) Deviation of $\chi(T)$ from the high temperature C-W law (orange dash-dot curve) described by a modified C-W law (eq.(1), red curve). 
b) Corresponding inverse susceptibility representations. Inset: High temperature inverse susceptibility compared with fitting functions. \label{F2}}
\end{center}
\end{figure}

Although the previous characterization of Eu$_2$Pd$_2$Sn \cite{ours} reveals an antiferromagnetic (AF) transition at moderated temperature $T_N=13.3$\,K, owing the large effective moment of Eu$^{2+}$ atoms: $\mu_{eff}$ = 7.93 $\mu_B$, significant magnetic correlations are triggered at higher temperature. 
In fact, pure paramagnetic Curie-Weiss (CW) behavior of the magnetic susceptibility: $\chi=C_c/(T-\theta)$, is only 
observed above about 70\,K. From that region a Curie constant  $C_c$($T \geq 70$\,K) =7.8\, emu\,K/Eu\,at.Oe was 
extracted together with a positive (FM) paramagnetic temperature $\theta_P = 18$\,K \cite{ours}, see the inset in Fig.~\ref{F2}b. The cusp at $T=T_N=13.3$\,K is the sign of the AF transition

However, a detailed analysis of the inverse susceptibility ($1/\chi$) reveals a slight positive 
curvature below about 60\,K. Although such curvature is frequently observed in systems exhibiting magnetic phenomena, like spin glass or Kondo effect, they are not expected to occur in this Eu$^{2+}$ compound. 
The periodic lattice distribution of magnetic atoms and their stable magnetic moment (with orbital moment $L=0$) prevent both scenarios.  

Since the previous measurements were performed 
in a field $B  = 1$\,T, in order to better investigate the origin of such deviation from the CW behavior further measurements at low field ($B=5$\,mT) were carried out. These results are presented in Fig.~\ref{F2}a 
and b in the range $T\leq 70$\,K revealing more details about that deviation.  

With the aim to search for the origin of such deviation we have fitted the experimental data using a modified CW-law function. The two constraints for that heuristic function are: to 
properly fit the deviation from the 
observed CW law and to reproduce it above 70\,K. Since the system is expected to continuously change from its  high temperature ($T\geq 70$\,K) paramagnetic configuration to another (at $T=T_N$), 
the most general formula is obtained by setting the two main parameters free to depend on  temperature as:
\begin{equation} \label{eq1}	
\chi(T)  = C_c(T)/[T-\theta_P(T)] + \chi_P
\end{equation}
The best fit was obtained using the following temperature dependencies: 
$C_c(T)=C_{HT}\,(1.8-e^{-\delta/T})$, with the factor  
$C_{HT}=7.87$\,emu\,K/Euat.Oe coincident with the Curie constant obtained at high temperature \cite{ours}, 
and  $\theta_P(T) = T+\theta_{LT}+ \theta_{fit}*e^{-\delta/T}$.
The extracted parameters for $\theta_P(T)$ are: $\delta =14$\,K, $\theta_{LT}= -4.5$\,K and $\theta_{fit}=21$\,K, see eq.(1) (red curve) in Fig.~\ref{F2}a and b. In the inset of 
Fig.~\ref{F2}b one can see how this fit fulfills the condition to follow the C-W law above 70\,K, after including a Pauli-like contribution: $\chi_P=2\, 10^{-3}$\,emu/Euat.Oe.

The physical meaning of the extracted parameters is the following. The Boltzman factor $e^{-\delta/T}$ represents the thermally driven access to the high temperature configuration described by the standard C-W law.
Interestingly, the 
characteristic energy of this promotion from LT to HT interaction configurations is $\delta = 14$\,K, very close to $T_N$. Although one cannot confuse a Boltzman-like energy promotion with a phase 
transition it is evident that, under cooling, the magnetic phase transition occurs once the LT configuration is sufficiently developed. 

Concerning the nature of the LT configuration, two features deserve to be highlighted. One is that $C_{LT}= 14.16$\,emu\,K/Euat.Oe nearly doubles the $C_{HT}$ value and the other is its negative value of $\theta_{LT}=- 4.5$\,K which indicates the dominant AF effect of involved interactions. The term: $\theta_{fit} *e^{-\delta/T}$, simply describes how the low temperature AF correlation are overtaken by 
the FM ones as temperature increases. To proceed to a more detailed comparison between HT and LT configurations, it is convenient to identify the involved interactions because they simultaneously contribute to the observed sign and value of $\theta_P(T)$.

Taking into account the local coordination of Eu atoms within the puckered-elongated hexagonal layers (see   Fig.~\ref{F1}a) one can see that the three involved intra-layer magnetic 
interactions ({$K_{ex}$}) can be described using the nomenclature proposed for a Kitaev type scenario \cite{Kitaev}. According to the scheme remarked in Fig.~\ref{F1}a by the magenta circle, $K_x$ and $K_y$ shall indicate the exchange interaction between Eu-NN belonging to the Eu zig-zag chains, while $K_z$ refers to 
the one along the elongate side of the hexagon. 

The fact that $\theta_{HT} = \theta_{LT} + \theta_{fit} = 16.5$\,K is positive and close to the value extracted from the standard C-W law, indicates the presence of dominant FM interactions between Eu-NN: $K_x$ and $K_y$. The $K_z$ interaction is expected not to be relevant because it acts on the Eu-NNN. As mentioned before, when $\theta_P(T)$ decreases with temperature it reveals 
that such FM interaction is compensated and overcome by an increasing AF interaction. Therefore, the observed  $\theta_P$ value corresponds to the additive criterion: $\theta_P \propto \Sigma_i K_i$ \cite{Blanco}, as a first approach for competing interactions. 

It is worth mentioning that the thermal promotion factor: $e^{-\delta/T}$, applied in Eq.(1) was inspired in the Bleaney-Bowers equation \cite{Bleaney} for the $\chi(T)$ dependence of a spin-dimer system composed by two nuclear 1/2-spins, where 
the energy gap of the resulting split singlet-triplet level scheme is temperature independent. That is a strong restriction because in the process to form a dimer there are two atomic GS (e.g. two s=1/2 doublets) that progressively transform into a singlet-triplet system without undergoing a first order transition, i.e. the involved levels change their relative energies driven by the arising magnetic 
interaction during the cooling process. Consequently in a real system the energy gap $\delta$ develops from zero till to stabilize once the dimer is formed. 
This is the reason to include the $e^{-\delta/T}$ factor in both temperature dependent parameters: $\theta_P(T)$ and $C_c(T)$. 
In this Eu based compound this process is certainly more complex because the starting GS is highly degenerat:  $N=2J+1=8$, due to the large $J=7/2$ value involved.

As mentioned before, applying Eq.(1) one extracts that $C_{LT}= 1.8*C_{HT}$. This is a relevant 
result because, according to the possibility of  Eu-dimers formation, one may analyze it in terms of the Curie constant of a dimer-quasiparticle: $C_{\rm D}$. In such a scenario, the unit of mass becomes the dimer, i.e. 2\,EuAt. Consequently, the dimer's unit of mass is '1\,mol' according to the formula 
unit Eu$_2$Pd$_2$Sn. Therefore, from $C_{LT} = 1.8 \times 7.87$\,emu\,K/EuAt.Oe one obtains $C_{\rm D}= 28.34$\,emu\,K/mol\,Oe.

The total angular momentum of each Eu-dimer $J_D$ depends on the different possible projections of the originary monents $J_{Eu}=7/2$. Starting with the maximum value $J_D=7$ one may evaluate the relative ratio between respective Curie constants $C_c \propto g_J^2 J(J+1)$ as:
\begin{equation} \label{eq2}	
 C_D/C_{HT}=  J_D (J_D +1)/ J_{Eu}(J_{Eu}+1) 
\end{equation} 
where the $g$-factor is considered not to change (i.e. the  angular moment remains $L=0$). Taking the maximum possible projection for $J_D$ the computed ratio between both parameters is: $C_D/C_{HT}= 
(7\times 8) / (3.5\times 4.5) = 56/15.75 = 3.55$. This ratio is quite close to the experimental one: 28.34/7.87 = 3.62. 
Notice that the chosen maximum possible projection: $J_D =7$ is the proper one because the following value: $J'_D = 6$, would given a much smaller ratio: 2.66. 

\begin{figure}
\begin{center}
\includegraphics[width=20pc]{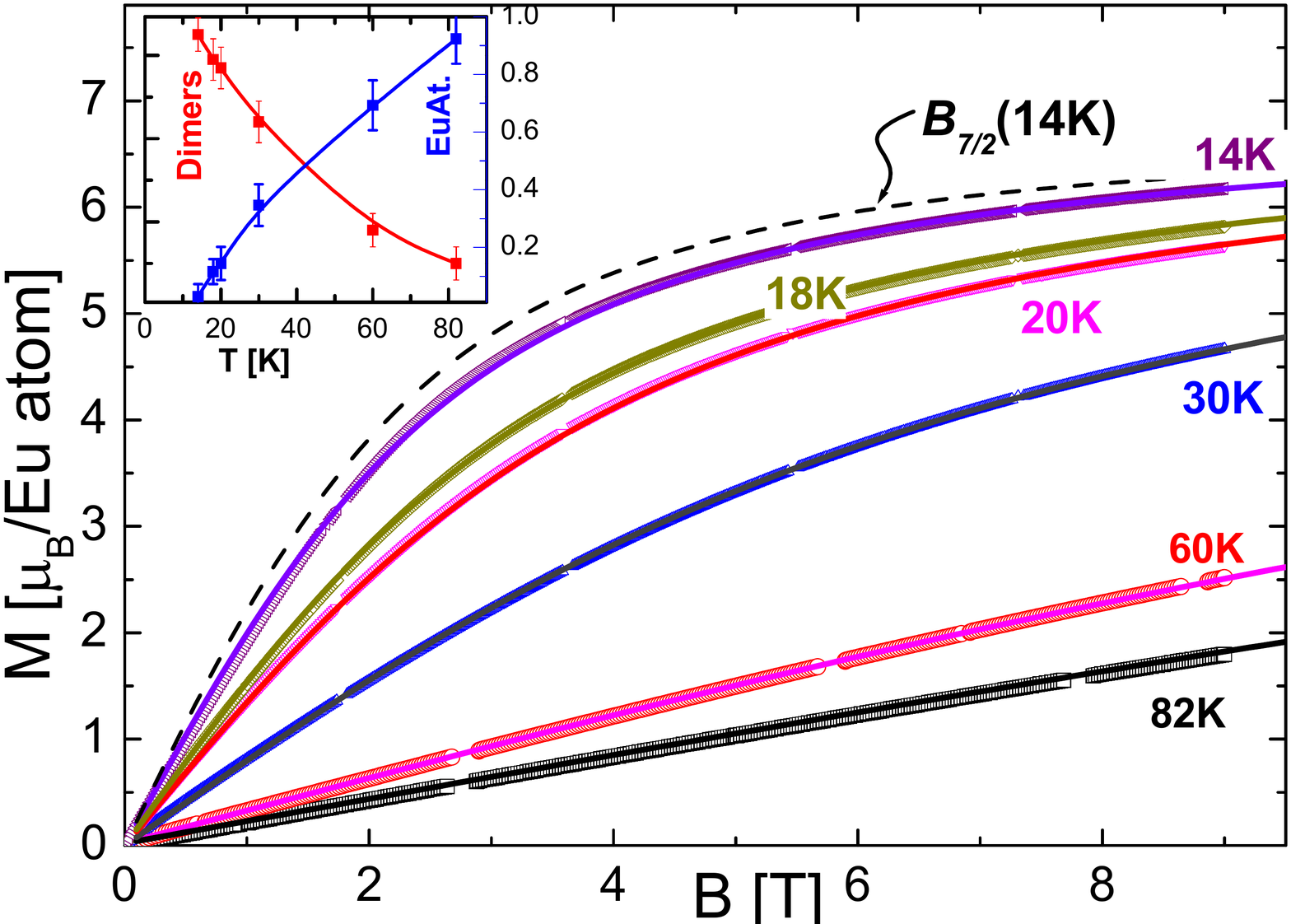}
\caption{(Color online)  Magnetization isotherms within the paramagnetic phase ($14 \leq T \leq 82$\,K). 
Continuous curves on each isotherm  represents the fit performed after having weighed $B_{7/2}$ and $B_7$ components (see the text). Black-dashed curve compares a 
pure Brillouin function computed with $J=7/2$ at 14\,K: $B_{7/2}(14K)$ with the experimental results (violet). Inset; red (blue) curve: fraction of dimers (single EuAt.) variation with decreasing (increasing) temperature. \label{F3}}
\end{center}
\end{figure}

In this scenario of Eu-dimers quasiparticles one has to consider that $K_{x}$ and $K_{y}$ interactions change their meaning because high temperature Eu-chains transform into dimer-chains. Consequently, approaching $T_N$ one may identify $K_{x}$ as the FM-intra-dimer interaction and $K_{y}$ as the AF-inter-dimer one. Therefore, from the observed $\theta_P(T<70\,K)$ evolution 
one learns that the AF-$K_{y}$ interaction becomes dominant because the FM-$K_{x}$ in the paramagnetic phase is becomes a sort of the dimer formation constituent. As it can be appreciated the mechanism of dimers formation requires further theoretic development.  

\subsubsection{Isothermal Magnetization above $T_N$} 

Similar phenomenology emerges in magnetization vs. filed studies within the same range of 
temperature. In Fig.~\ref{F3}, isothermal magnetization curves measured between $\approx T_N$ (i.e.  14K) and 82K are shown. The experimental curves can be properly fitted at 82 and 60K 
with the paramagnetic Brillouin function: 
\begin{equation} \label{eq.(3)}	
B_J(x)=(a/b)/\tanh[(a/b) x]-(1/b)/\tanh[x/b]
\end{equation}
where $a=(2J+1)$, $b=2J$ and 
$x = gJ_{Eu}\mu_B B/k_B T = 4.66\times (B/T)$ using the $g_J=2$ and $J_{Eu} =7/2$ values. 

However, using this $J_{Eu}$ factor in Eq.(3), the computed curve for lower tempertures isotherms progressively depart from the measured magnetization. In Fig.~\ref{F3} the $M(B)$ results at $T = 14\,K$ (violet points) are 
compared with the computed (dashed) curve labeled $B_{7/2}(14\,K)$.
Alternatively, a better fit 
of $M(B)$ at that temperature is obtained applying the same function but with a $J_D= 7$, i.e. $B_{7}(y)$ 
and therefore $y = gJ_D\mu_B  B/k_B T = 11.6(B/T)$ which corresponds to the presence of Eu-dimers.

\begin{figure}
\begin{center}
\includegraphics[width=20pc]{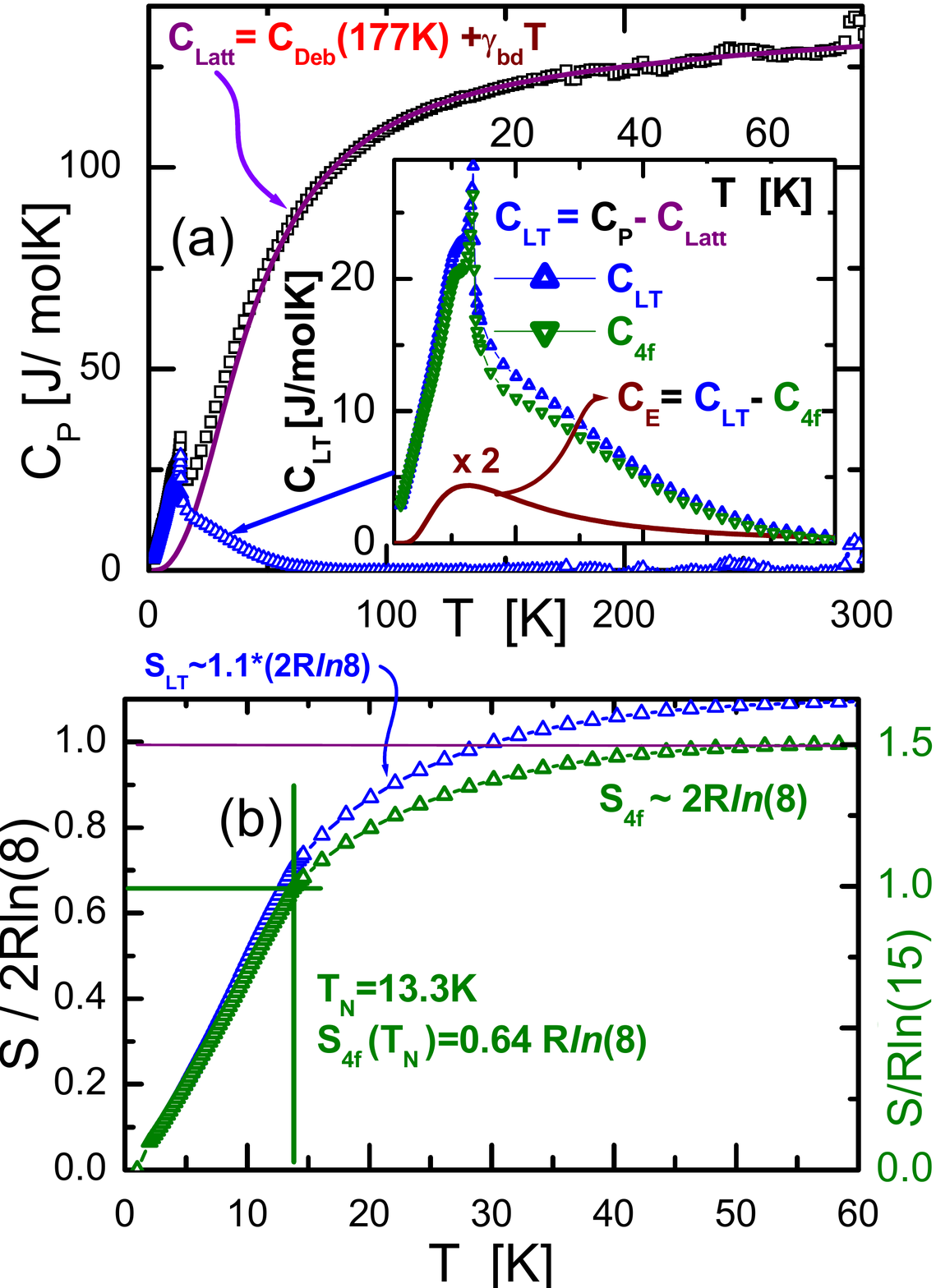}
\caption{(Color online) a) Measured specific heat up to room temperature described by a Debye function (with $\theta_{Deb}=177$\,K) and an electronic-band $\gamma_{bd}*T$ contribution (see the text). 
Inset: Low temperature specific heat contribution ($C_{LT}
$: blue points) after phonon and electron band subtraction, and an Einstein type soft phonon contribution: $C_E$ (two times for clarity, brown curve). b) Entropy evolution 
before ($S_{LT}$: blue points)  and after ($S_{4f}$: green poins) the soft phonon contribution subtraction. \label{F4}}
\end{center}
\end{figure}

This difference allows to evaluate the dimer’s fraction variation as a function of temperature. 
At intermediate temperature: $ 82 \geq T \geq 14$\,K, both functions $B_{7/2}(x)$ and $B_7(y)$ are applied to fit the measured magnetization as follows: $M = M_{Sat} [E\times B_{7/2}(x) + 
D\times B_7(y)]$, where the weight factors represent 
respective contributions of single Eu atoms (E) and Eu-Eu dimers (D) with the condition: $E + D = 1$, and being $M_{Sat}
$ the saturation magnetization: $M_{Sat}(2\,K,9\,T) = 6.85 \mu_B$ \cite{ours}. 
According to $M(B)$ results, dimer’s condensation starts around 80\,K and concludes at $T_N = 13.3$\,K, see inset in Fig.~\ref{F3}.
This process begins at a temperature above which the extrapolated $\theta_P$ = 19\,K reveals the presence of a dominant FM exchange interactions: $K_{x,y} >$ 0 between Eu-NN and ends at 
$T_N$ where $K_{x} \neq K_{y}$

These fits confirm that, while the FM-$K_x$ interaction yields the condensation of Eu-dimers, the increasing AF-inter-dimer $K_y$ progressively overcomes it producing the observed thermal 
dependence of $ \theta_P(T)$. 

\subsubsection{Specific Heat and Entropy}

The temperature dependence of specific heat $C_P(T)$ of Eu$_2$Pd$_2$Sn was measured from 2\,K up to room temperature. 
Above about 60\,K the lattice contribution ($C_{Latt}$) is notably well fitted accounting for phonon ($C_{Deb}$) and band electrons ($\gamma_{bd}*T$) components only: $C_{Latt} = C_{Deb} +\gamma_{bd}*T$.
Respective contributions are described by a Debye function with a Debye temperature: $\theta_{Deb} = 177$K and  $\gamma_{bd}= 25$mJ/molK$^2$ typical for band electrons, see Fig.~\ref{F4}a. 

At lower temperature ($T < 60$\,K) localized $4f$ electrons of Eu atoms start to contribute: $C_{LT} = C_P – 
C_{Latt}$. However, the corresponding 
entropy evaluation up to about 60\,K:  $S_{LT}=  \int_{0K}^{60} C_{LT} /T dT$ , exceeds the expected value for $(2J+1)$ degeneracy of Eu$^{2+}$ GS: 2\,R$\ln(8)$, by about 10$\%$, see blue points in Fig.~\ref{F4}b, where R is the gas constant.
Such excess of entropy can be accounted by an Einstein-type contribution \cite{Einst} ($C_E$) from the difference between $C_{LT}$ and $C_{4f}$, see the inset of Fig.~\ref{F4}a.
\begin{equation} \label{eq5}
C_E= \rm {R} (2/5) \omega  (\Delta/T)^2 e^{\Delta /T}/ [1+\omega * e^{\Delta /T}]^ 2 
\end{equation}
This contribution appears as a soft phonon excitation with $\Delta =30$\,K.

As a thermodynamic parameter, specific heat does not provide direct microscopic information concerning the origin of such soft phonon. Nevertheless, the (2/5) pre-factor indicates that this excitation only involves two of the five atoms of the formula unit, e.g. the two Eu or two Sn. 

A complementary information is provided by the fact that the factor $\omega$ indicates a double degeneracy of the excited 
level because $\omega  =1/2$ \cite{W,Gopal}. 
This distribution of the levels degeneracy supports the dimers formation 
because at low temperature there is only one possible configuration whereas the excited state has two possibilities. Starting by the high temperature configuration one observes that 'right'  
and `left' Eu-Eu interactions within the Eu chains have the same energy since $K_{x} = K_{y}$. However, once the dimer condenses there is only one remnant configuration left  
because one of those interactions (previously identified as $K_{x}>0$) has driven the 
quasiparticle formation. The other ($K_{y}$) plays the role of the inter-dimers interaction along the chains. A schematic representation of this $T\to T_N$ configuration is depicted in Fig.~\ref{F11}. This description is in agreement with the temperature dependence of $
\theta_P(T)$ because its high temperature ($T\geq 70$\,K) positive value reflects the FM-$K_{x,y}>0$ of all Eu-NN interactions, whereas the slightly negative value at $T\to T_N$ is 
the result of the arising inter-dimers AF-$K_{y}<0$ interaction. 

Such a strong modification of both $K_{x,y}$ interactions may trigger the claimed soft phonon as a sort of magnetoelastic effect reflected in a slight displacement of alternated Eu. In fact  that the symmetric ('right' and 'left') inter-atomic bonds along the chain above 70\,K is replaced by one $intra$-dimer bond in one direction of the chain and one $inter$-dimer interaction in the oposite one. \\

That soft phonon contribution is subtracted to obtain the pure $4f$ electronic contribution: $C_{4f} = C_{LT} - C_E$, see the green points in the inset of Fig.~\ref{F4}a, whose associated entropy is the expected value: $S_{4f}$=2\,R$\ln8$ shown in Fig.~\ref{F4}b. 
In that temperature evolution of the entropy, one can see that the value at $T = T_N = 13.3$\,K is: $S_{4f}(T_N) = 0.64\times 2$\,R$\ln8$. 
Notably, this value equals  R$\ln15$, which corresponds to the entropy of a dimer with total angular momentum: $J_D=7$.
Consequently, the remnant entropy gain between $T_N$ and $T\approx 50$\,K: 2\,R\,$\ln8-$R$\ln15$, is the entropy 
condensed along the Eu-dimers formation process. Notice that in this entropic description the unit of mass is the 'mol', that means 2\,EuAt. at high tempertures and 1\,dimer approaching $T_N$.  

The specific heat provides relevant information upon the phase transition through the jump \cite{jump} depicted in the inset of Fig.~\ref{F4}a. 
For the case of $J_{Eu} = 7/2$: $\Delta C_{4f}(T_N)$ = 20.1/Euat.K and for $J_D = 7$ is only slightly increased because $\Delta C$ tends to saturate for high $J$ evalues. 
However, because nearly 1/3 of the degrees of freedom are already condensed at $T_N$ ($S_{4f} (T_N)= 0.64\times 2\,R\ln8$), the expected $\Delta C_{4f}(T_N)$  jump is $\approx 13$\,J/Euat.K $\approx 26$\,J/mol K, very close to the value observed experimentally. 

The formation of Eu-dimer quasiparticles is therefore the most relevant message obtained from the analysis of the magnetic and thermal properties of the paramagnetic phase of this compound. This finding is in agreement with 
the exceptionally large Eu-Eu electronic overlap mentioned in the {\bf Introduction} after considering the reduced Eu-NN  
spacing within the chains. 

\subsection{Magnetically Ordered Phase}
\subsubsection{Magnetic susceptibillity}

\begin{figure}
\begin{center}
\includegraphics[width=20pc]{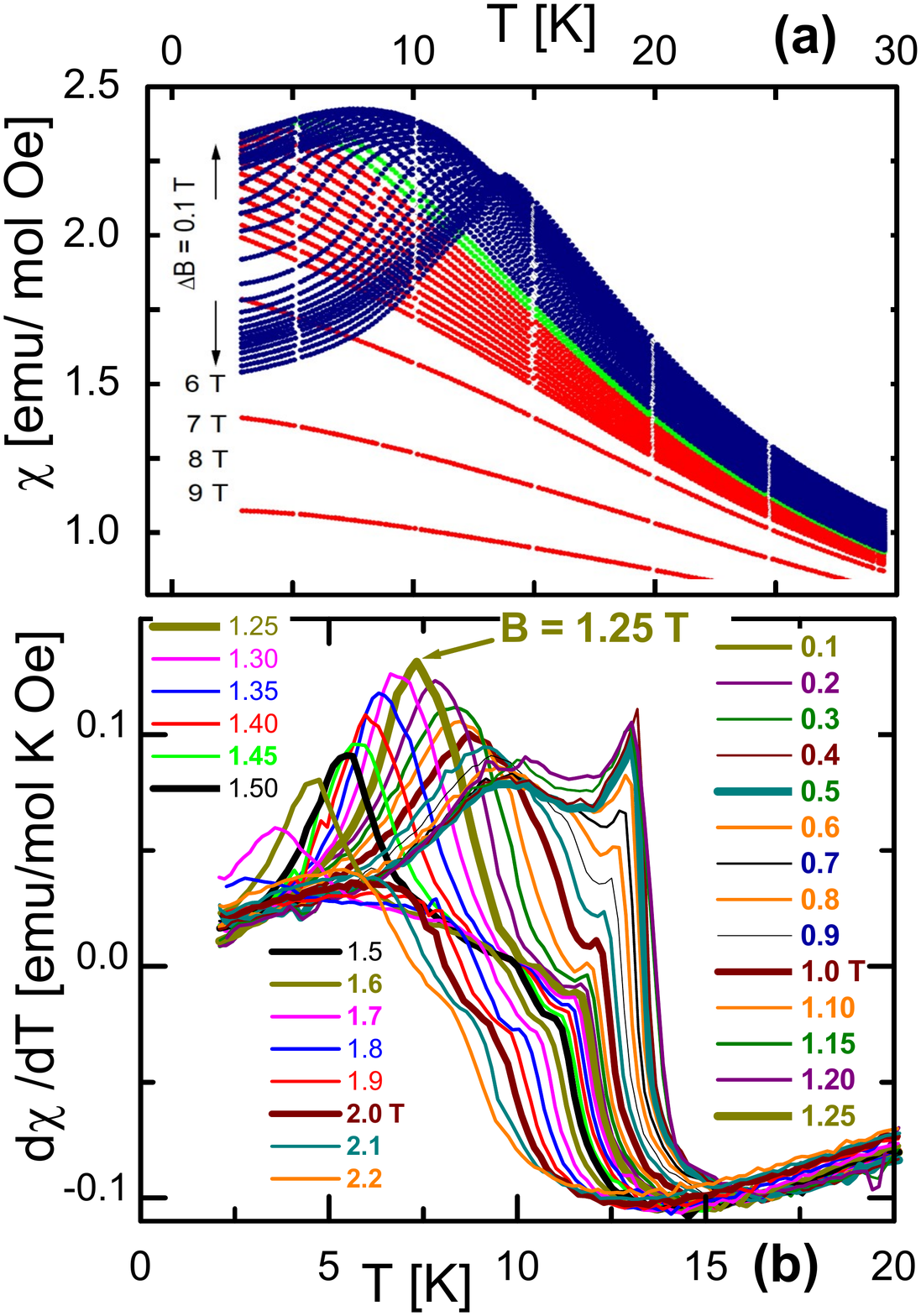}
\caption{a) (Color online) Detailed isodynas at constant applied field of magnetic susceptibility as a function of temperature up to 30\,K, with magnetic field sweeping between $0 \leq B \leq 6$\,T with 0.1T increase.  b) Temperature derivative of those curves up to 20\,K. \label{F5}}
\end{center}
\end{figure}

The magnetic susceptibillity $\chi(T)$ was measured in detail below 30\,K, increasing the magnetic field in steps of 0.1\,T, see  Fig.~\ref{F5}a. At 
low fields a clear maximum at $T =T_N =13.3$\,K  is observed, followed by a weak shoulder at 10K. This pattern holds up to $\approx 0.5$\,T where the relative intensity of both features progressively changes. 

\begin{figure}
\begin{center}
\includegraphics[width=20pc]{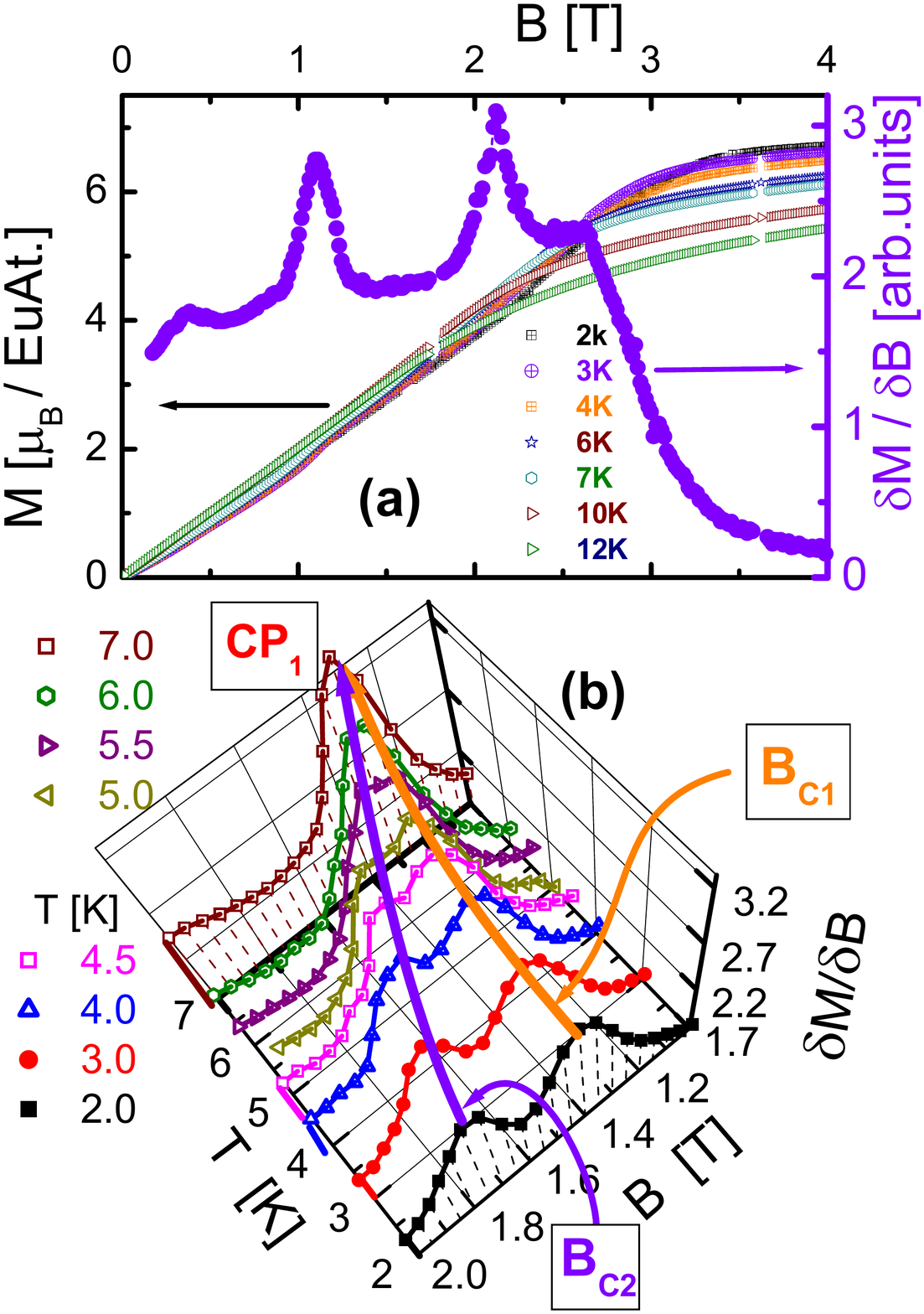}
\caption{(Color online) a) Left axis: $M(B)$ results within the ordered phase ($2 \leq T \leq 12$\,K) measured up to $B = 4$\,T. Right axis: example 
of the $\partial M/\partial B$ derivative at $T=2$\,K.  
b) 3D representation of respective $\partial M/\partial B$ derivatives between $2 \leq T \leq 7$\,K and $1.1 \leq B \leq 2$\,T 
ranges. Curves labeled as $B_{c1}$ (orange) and $B_{c2}$ (violet) follow respective maxima of the $\partial M/\partial B$ peaks 
showing their convergence into a critical point CP$_1$ at $T_1=7$\,K and $B_1 =1.25$\,T. \label{F6}}
\end{center}
\end{figure}

In order to make more evident the effect of the magnetic field, the temperature derivative: $\partial \chi/\partial T$, was computed in the range: $0 \leq B \leq 2$T 
as depicted in Fig.~\ref{F5}b. In that figure one can see how the $\partial \chi(T_N)/\partial T$ peak decreases in intensity whereas the dome at 10\,K starts to increase for $B \geq 0.6$\,T while it shifts towards lower temperature. 

Above $B \approx 1$\,T, the susceptibility reveals further modifications in the magnetic behavior.
Increasing the field, the $\partial \chi/\partial T$ dome sharpens showing a maximum intensity at: $B_{cr} =1.25$\,T and  $T_{cr} = 7.3$\,K, 
see Fig.~\ref{F5}b. Above that critical field the $\partial \chi /\partial T$ dome decreases in 
temperature and intensity till to vanish around $B = 2$\,T.

\subsubsection{Magnetization}

According to the results presented in ref \cite{ours}, in the magnetically ordered phase $M(B)$ increases nearly but not strictly linearly between $0 \leq B \leq 2.5$T, see Fig.~\ref{F6}a (left axis). Then, above $B\approx 3$\,T the magnetization tends to saturate and for the lowest measured temperature it reaches the  $M_{sat}(2\,K) = 6.85\mu_B$/EuAt value.

Since the investigated samples are polycrystalline, eventual anisotropic effects can be highlighted analyzing the $\partial M/\partial B$ derivative. In the case of linear $M(B)$ contribution in one (or two) 
directions their respective derivatives shall be field independent and therefore this procedure allows to reveal details of field dependence on the third one. In this case, the changes in the $M(B)$ 
slopes are well identified as it is shown in Fig.~\ref{F6}a (right axis) for the $T=2$\,K curve. The two 
well defined maxima observed sign the changes of slope $M(B)$, indicating the presence of anisotropy in at least one of the axes.

\begin{figure}
\begin{center}
\includegraphics[width=20pc]{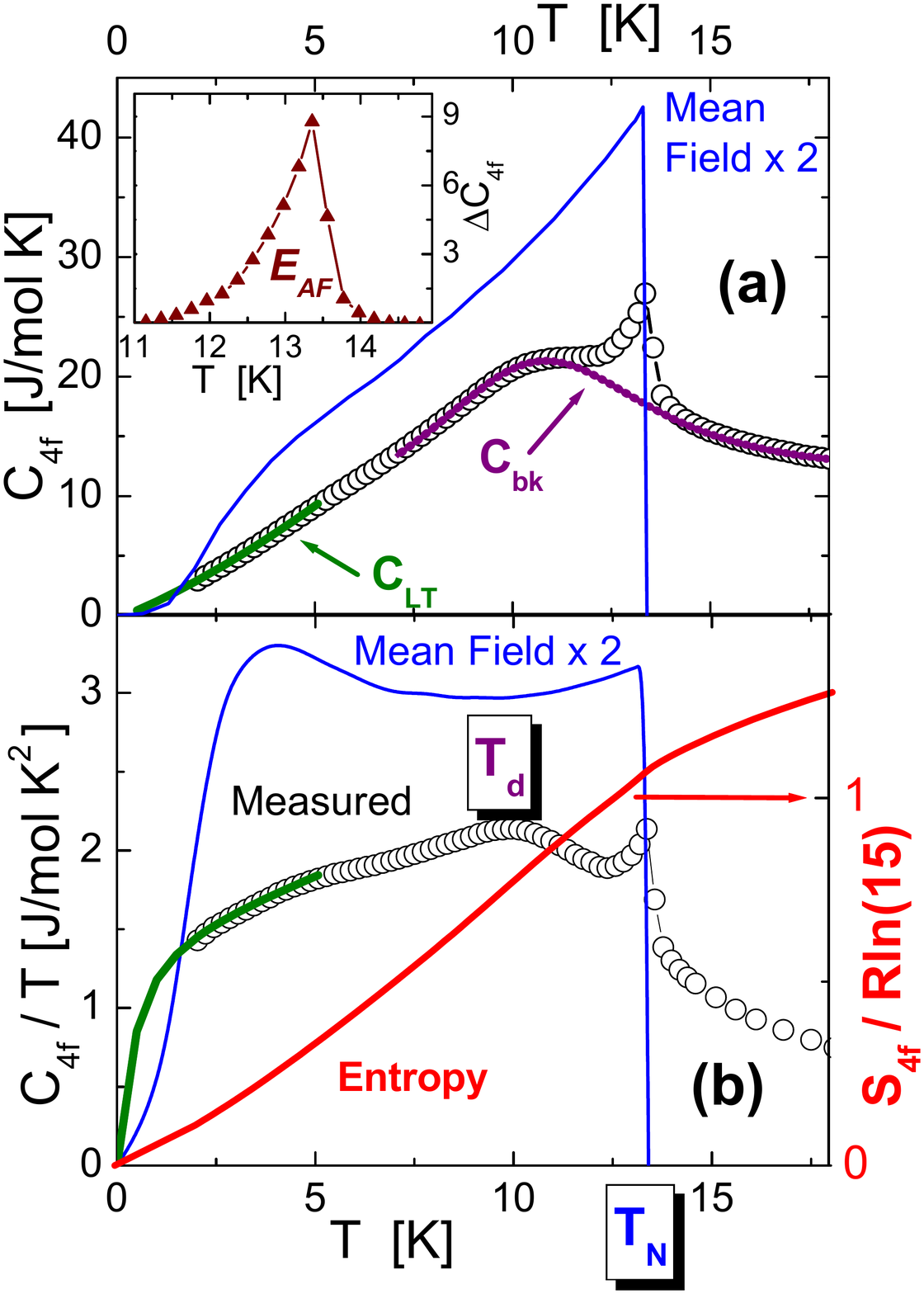}
\caption{(Color online) a) Detail of the $4f$ contribution to specific heat in the ordered phase. Green curve ($C_{LT}$): low temperature fit highlighting quasi linear temperature dependence with small gap at $T\to 0$. Purple curve: $7\leq T \leq 20$\,K background ($C_{bk}$) fit to evaluate the enthalpy $E_{AF}$ involved in the transition. Blue curve: mean field prediction for a $J=7/2$ system \cite{MF}. 
Notice that the factor 'x2' is included because of the mol = 2EuAt units. Inset: $\Delta C_{4f}$ curve after $C_{bk}$ subtraction, see the text. \\
b) Left axis: measured specific heat divided temperature compared with mean field prediction (blue curve) for $J=7/2$. Green curve: $C_{LT}/T$ representation of the low 
temperature fit remarking the presence of a $T\to 0$ gap. Right axis: (red curve) temperature dependence of the entropy normalized to Rln(15). 
\label{F7}}
\end{center}
\end{figure}

In Fig.~\ref{F6}b the $\partial M/\partial B$  isotherms are presented in a three dimension (3D) representation within the $2 \leq  T \leq 7$K and $1.1 \leq B \leq 2$T ranges. There, one can appreciate the evolution of two 
characteristic maxima related to respective critical fields:
$B_{C1}$ running from $B = 1.4$\,T at 2\,K to $B = 1.25$\,T at 7\,K, and $B_{C2}$ running from $B = 1.8$\,T at 2\,K to $B = 1.25$\,T at 7\,K.
Both curves merge at a critical point (CP$_1$) at $B_{cr1} = 1.25$\,T and 7\,K. Under further increase of the temperature, a single curve follows up to $B = 0.9$\,T at 11\,K (not shown for clarity).
Another critical curve, $B_{C3}$ , runs from $B = 2.45$\,T at 2\,K, to $B = 1.15$\,T at 12\,K, not included within the range of this figure.

\begin{figure}
\begin{center}
\includegraphics[width=20pc]{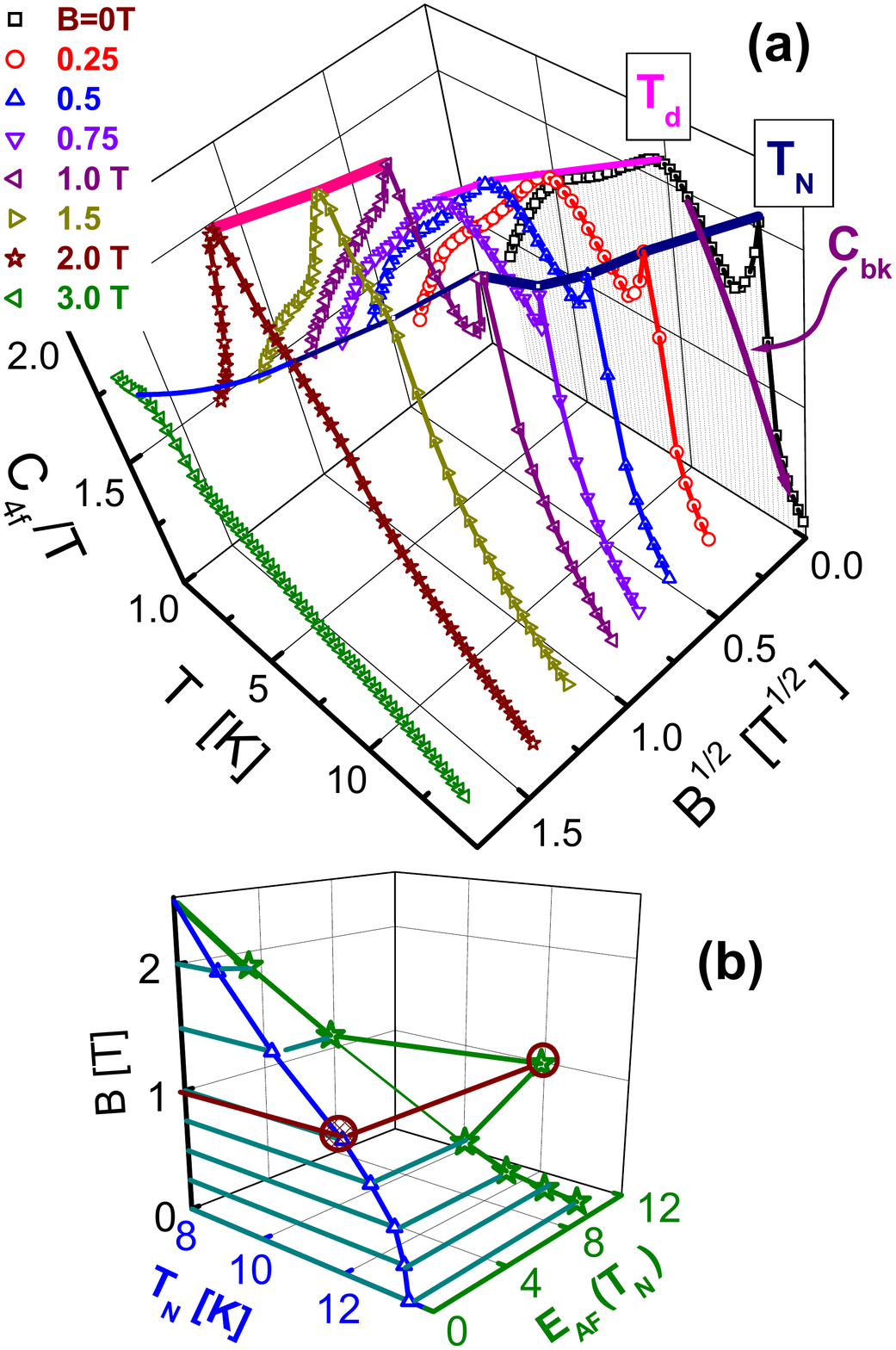}
\caption{(Color online) a) 3D representation of temperature and field dependence of $4f$ electrons contribution 
to specific heat, remarking the AF transition $T_N(B)$ (blue curve) and the dome $T_d(B)$ (magenta and pink curves) variations. The purple curve at $B=0$ represents the $C_{bk}(T)$ fit from  Fig.~\ref{F7}a. Notice that the magnetic field axis is defined as the square root $B^{1/2}$ for a clearer distribution of the curves. 
b) 3D representation of the enthalpy $E_{AF}$  associated to the $T_N(B)$ transition (see the text) showing the strong increase at $B =1$T (brown lines). \label{F8}}
\end{center}
\end{figure}

\subsubsection{Specific heat}

Within the ordered phases, the specific heat of magnetic elements with large $J > 3/2$ values show a clear shoulder at about $1/3$ of $T_N$, which is originated in the increasing Zeeman splitting of the manifold GS levels of the paramagnetic state driven by the arising 
internal molecular field. This effect is particulary clear when there is not CEF effect present \cite{MF} and therefore the most prominent examples for this scenario are Gd$^ {3+}$ and Eu$^ {2+}$ based compounds with $J=7/2$, see for example Refs. \cite{Kumar10,MaruyaGe3,EuNi5As3}. 

In the case of Eu$_2$Pd$_2$Sn, however, such shoulder is extremely weak as it can be appreciate in Fig.~\ref{F7}a in comparison with that of the mean field (MF) prediction (blue curve). In this case it can be visualized around 4.5\,K only in a $C_{4f}/T$ representation, see Fig.~\ref{F7}b. 
Furthermore, at first glance the $C_{4f}(T)$ dependence looks quite linear below about 7.5\, K suggesting a  continuous spectrum of excitations instead of the Zeemann like distribution 
of levels. For a more detailed analysis, we have fitted the low temperature ($T < 5$\,K) range with the  function: $C_{LT}= \alpha T \exp(-d/T) + \beta T^2$, see green curve in Fig.~\ref{F7}a. The large coefficient $\alpha = 
1.5$\,J/mol\,K$^2$ reveals a high density of excitations, only limited at low energy by a small gap of anisotropy $d=0.3$\,K. The term $\beta T^2$, with $\beta = 0.09$\,J/mol\,K$^2$, is a correction to account for the onset of the slight hump around 4.5\,K. These parameters are compatible with a low dimensional AF order \cite{Gopal}, including helical order \cite{EuCo2P2}. 

On the other hand, measured $C_{4f}(T)$ exhibits a well defined dome at $T_d = 10$\,K, which is not predicted by the MF description but for modulated magnetic structures which depend on the relative magnitude of the 
involved exchange couplings \cite{Blanco}. 

Concerning the field dependence, the $4f$ electronic contribution to specific heat $C_{4f}(T)$ in different applied fields was previously published in ref \cite{ours}. Those results are collected in a 3D representation of $C_{4f}/T$ in  Fig.~\ref{F8}a, where  the AF transition $T_N(B)$ and the temperature of the dome, $T_d(B)$, are traced as a function of magnetic  field.  
 
In the figure the sharp peak of $C_{4f}(T_N)/T$ is observed up to $B =1$\,T. Beyond that field it transforms 
into a shoulder. With the aim to investigate whether the sudden increase of that peak at $T_N$(12\,K,1\,T) has the character of a 
critical point, we have evaluated the related enthalpy, $E_{AF}(T_N ,B)$, by subtracting a background specific heat $C_{bk}(T,B)$ to the measured $C_{4f}(T)$. Such a background is defined by a continuous curve which 
fits \cite{bkgr} the measured $C_{4f}(T)$ above and below $T_N$ within a window of $\pm 10\%$, see the 
purple $C_{bk}(T)$ curve in Fig.~\ref{F7}a and in Fig.~\ref{F8}a for the case of the $C_{4f}(T,0)$ results. 

The extracted enthalpy of the transition is computed as  $E_{AF}= \int(C_{4f} - C_{bk}) dT$ and depicted in  
Fig.~\ref{F8}b as a function of field and $T_N(B)$. One can appreciate how $E_{AF}(T_N,B)$ decreases monotonously from about 8\,J/mol at 
$B=0$ to zero at $B\approx 2.5$\,T. However, at $B=1$\,T a drastic increase up to $E_{AF}=11.6$\,J/mol is observed indicating the presence of a 
singularity likely related to another Critical Point CP$_2$ at $T_N$(12\,K,1\,T).

Turning back to Fig.~\ref{F8}a, one can see that at that as a function of field the shoulder identified as $T_d(B<1T)$ (magenta curve) turns into a sharp peak, which holds up to $B=2$ \,T (pink curve). 

\begin{figure}
\begin{center}
\includegraphics[width=20pc]{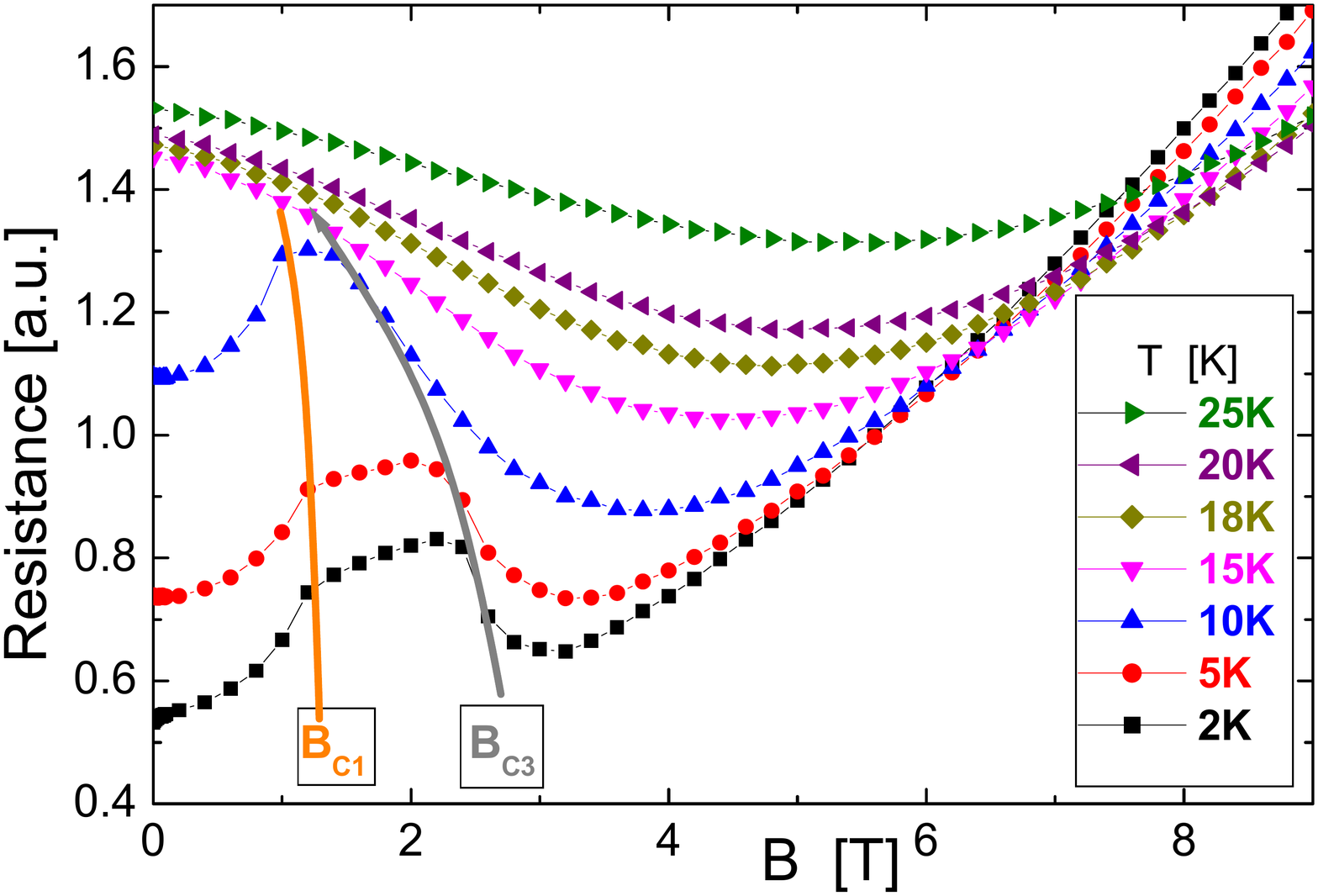}
\caption{(Color online) Magnetoresisence at different temperatures up to $B=9$\,T, showing how critical fields $B_{C1}$ and $B_{C3}$ converge between 10 and 15\,K around 1.2\,T. \label{F9}}
\end{center}
\end{figure}

\subsubsection{Magnetoresistence}

In Fig.~\ref{F9}, the magnetoresistence (MR) clearly exhibits two different regimes, above and below $T_N$. In the paramagnetic one (i.e. $T \geq T_N$) a broad minimum centered around $B=5$\,T indicates the scattering 
with short range interactions that are progressively suppressed by field. These magnetic correlations can be associated to the competition between FM-intra-dimer and AF-inter-dimer interactions as $T\rightarrow T_N$. 

Such behavior is strongly modified within the magnetically ordered phase, where a clear 
plateau starts to develop at around 10\,K that expands by decreasing temperature down to $T =$ 5 and 2\,K. 
The borders of that plateau can be associated to two critical fields: $B_{C1}$ and $B_{C3}$, which converge to the already identified critical point CP$_2$ at $T_{cr2}= 12$\,K and $B_{cr2} = 1.15$\,T. In these series of measurements the critical field $B_{C2}$ cannot be distinguished from $B_{C1}$. 

\section{Discussion}

\subsection{Magnetic Phase Diagram}

All the experimental information on magnetic, thermal and transport properties collected from Eu$_2$Pd$_2$Sn allows to draw a magnetic phase diagram presented in Fig.~\ref{F10}. Starting from the low 
temperature region, two critical fields: 
$B_{C1}$  and $B_{C2}$, are well described by respective peaks in $\partial M/\partial B$ and related maxima in $\partial \chi/\partial T$ derivatives. 
Both phase  boundaries converge on a critical point CP$_1$ at $T_{cr1}=7$\,K and $B_{cr1} = 1.25$\,T. Other two lines end on that CP$_1$, one related to the $T_d(T,B)$ dome observed in specific heat and $\partial \chi/\partial T$,  
and the other to a kink in $\partial \chi/\partial T$. 

The second critical point was identified through the sudden increase of the enthalpy in the $T_N(B)$ transition presented in Fig.~\ref{F7}b. Another critical field curve, related to an edge in the $\partial M/\partial B$ slope: $B_{C3}$, runs from $B= 2.45$\,T at $T=2$\,K and converge on this critical point CP$_2$ at $T_{cr2}=12$\,K and $B_{cr2} = 1.15$\,T.  
At higher fields, a sort of slit in $\partial \chi/\partial T$ indicates the limit of the region where the fully polarized (FM-FP) paramagnetic moments begins, which vanishes around 4\,K.    

\subsection{Searching for the identity of each magnetic phase}

Although the different phase boundaries determined through the present thermodynamic and transport results  are quite well defined, the polycrystalline nature of the sample doodles any direct information about the anisotropy of 
their magnetic structures. This handicap can be relieved by analyzing the anisotropic effects in systems with very similar behavior. One of the most appropriate cases is EuNiGe$_3$, studied on single crystals \cite{RyanGe3, MaruyaGe3}, 
which exhibits significant coincidences like: i) it crystallizes in non-centrosymmetric structure, ii) it orders AF at the same temperature ($T_N = 13.2$\,K), iii) that transition is 
followed by a broad hump at $T_d = 10.5$\,K, and iv) it shows equivalent jumps in magnetoresitance. All these likeness converge into a very similar phase diagram. 
Notably, the step like increase of the magnetization clearly observed in EuNiGe$_3$ \cite{MaruyaGe3} is in full coincidence with that obtained from the $\partial M/\partial B$ of Eu$_2$Pd$_2$Sn. From this feature a relevant 
information can be extracted from the 1.8\,K isotherm $M(B)$ of EuNiGe$_3$. The strong 
anisotropy is characterized  by a staircase increase on the easy axis of magnetization [001], whereas on the ‘ab’ plane $M(B)$ increases quite linearly up to its saturation value $M_{sat}=7\mu_B$/EuAt. 
In polycrystalline Eu$_2$Pd$_2$Sn such staircase increase is certainly not observed because of the random distribution of the crystals, however the $\partial M/\partial B$ derivative (see  Fig.~\ref{F6}a) transforms the 
linear $M(B)$ components into a constants, highlighting the blurred steps as well defined peaks. Since only two 
peaks are clearly seen in Eu$_2$Pd$_2$Sn polycrystalline samples, it strongly suggests that only one direction 
of magnetization is responsible for those discontinuities.

\begin{figure}
\begin{center}
\includegraphics[width=20pc]{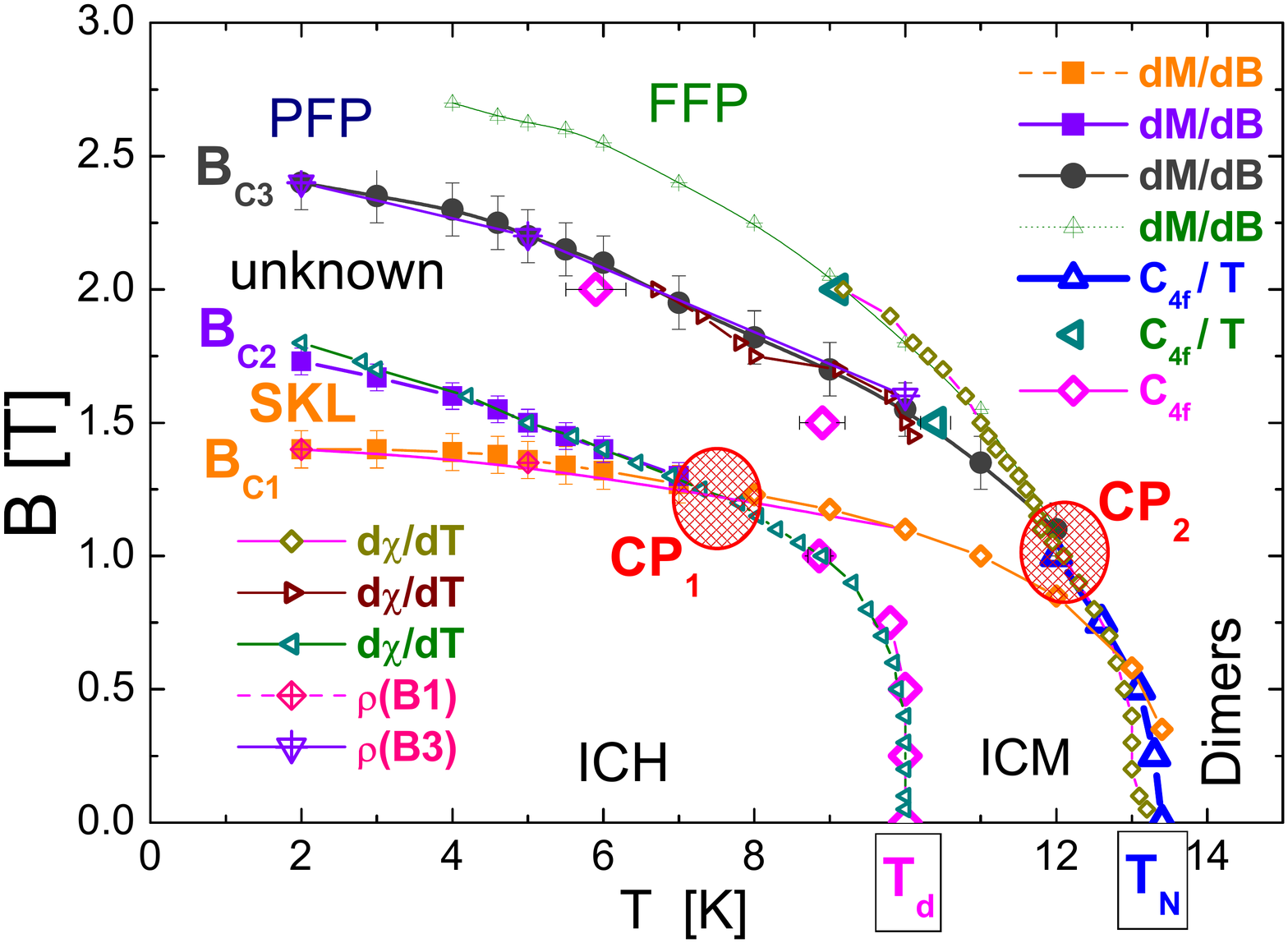}
\caption{(Color online) Magnetic phase diagram of Eu$_2$Pd$_2$Sn where different phase boundaries are defined by respective anomalies in magnetic, thermal and transport properties. Three main critical fields are included: 
$B_{C1}$ and $B_{C2}$, that converge in a critical point CP$_1$(7\,K,1.25\,T), and  $B_{C3}$ that ends on the critical point CP$_2$(12\,K,1.15\,T). Two phase boundaries arise from $B=0$: $T_d(B)$ indicated by the dome traced in Fig.~\ref{F8}a and by the AF transition $T_N(B)$. 
The proposed magnetic structures of the different phases are: ICM (Incommensurate Modulated \cite{RyanGe3}), ICH (Incommensurate Helicoidal), SKL (Skyrmion Lattice \cite{Xtian}), PFP (Partially Field Polarized) and FFP (Fully Field Polarized). \label{F10}}
\end{center}
\end{figure}

Concerning the specific heat properties, apart from the mentioned similarities in the jump at $T_N$ and the dome at $T_d$, there is a clear difference in the tail of $C_{4f}(T)/T$ right above $T_N$. This difference can 
be understood by the fact that, whereas in EuNiGe$_3$ the standard magnetic correlations develop as precursors of the magnetic transition, in Eu$_2$Pd$_2$Sn there is a dimers condensation which dominates the scenario 
above $T_N$. In the former compound the 
typical positive curvature $C_{4f} \propto T^{-2 }$ is observed (see also \cite{Blanco,EuNi5As3,EuIr2P2,GoetschGe3}), while in the latter there is a nearly linear increase approaching $T_N$, see the inset in 
Fig.~\ref{F4}a.
The consequence of this different condensation of degrees of freedom above $T_N$ is reflected in respective values of the entropy at $T_N$: in 
EuNiGe$_3$ \cite{MaruyaGe3} one has $S_{4f}(T_N)=0.82Rln(8)$, with the remaining entropy reached at ~30\,K, while in the case of Eu$_2$Pd$_2$Sn it is $S_{4f}(T_N) = 0.64Rln(8)$, with the remaining 
entropy reached at ~50\,K as indicated in Fig.~\ref{F4}b. 
Strictly, this large amount of involved entropy above $T_N$ is a 'necessary' condition for dimers formation, otherwise there would not be any condensation.
This difference is obviously reflected in the $C_{4f}(T<T_N)$ intensity despite their temperature dependencies look very similar. 

The large difference in the development of short range correlations above $T_N$ between these compounds is consistent with the difference of respective FM  paramagnetic temperatures: $\theta_p = 5$\,K for EuNiGe$_3$ 
and 19\,K for Eu$_2$Pd$_2$Sn, and remarks the presence of strong Eu-Eu interaction as responsible for the dimers formation in the latter compound.

Further similarities between these compounds are observed in magnetoresistence (MR) as well, being the most relevant a sharp plateau in MR($B$). Interestingly, this plateau is observed in EuNiGe$_3$ with the field applied on the [001] direction only, in agreement with the features detected in $M(B)$. 
Coincidentally, the magnetic phase diagram of EuNiGe$_3$ single crystals with field in the $B//[001]$ direction also resembles the one of Eu$_2$Pd$_2$Sn polycrystals, including the critical points identified in Fig.~\ref{F10}. 
Last but not least, the magnetic phase diagram of EuNiG$_3$ with $B$ applied in the [100] direction may explain the presence of the uppermost (vanishing) curve between partially and fully field polarized spins (PFP and FFP) phases.

All these similarities support the possibility to replicate the different magnetic structures determined in EuNiGe$_3$ into the Eu$_2$Pd$_2$Sn phase diagram. For example, some spectrosopic studies performed in the 
former compound, like $^{151}$Eu Mössbauer spectra \cite{MaruyaGe3}, provides insight for describing possible scenarios. From those results it is concluded that the 13.2 K transition leads to an Incommensurate (ICM) AF intermediate phase, followed by a transition near 10.5 \,K into a commensurate AF configuration. 
Latter studies \cite{RyanGe3} indicate that this compound adopts a complex Incommensurate Helicoidal (ICH) at low temperature which transforms into an Incommensurate Sinusoidal modulate up to $T_N$. 
Notably, from these measurements this transition does not appear to be a conventional second order phase transition, rather a collapse of the long-range order \cite{RyanGe3}.

\subsection{Magnetic structure reformulation}

\begin{figure}
\begin{center}
\includegraphics[width=22pc]{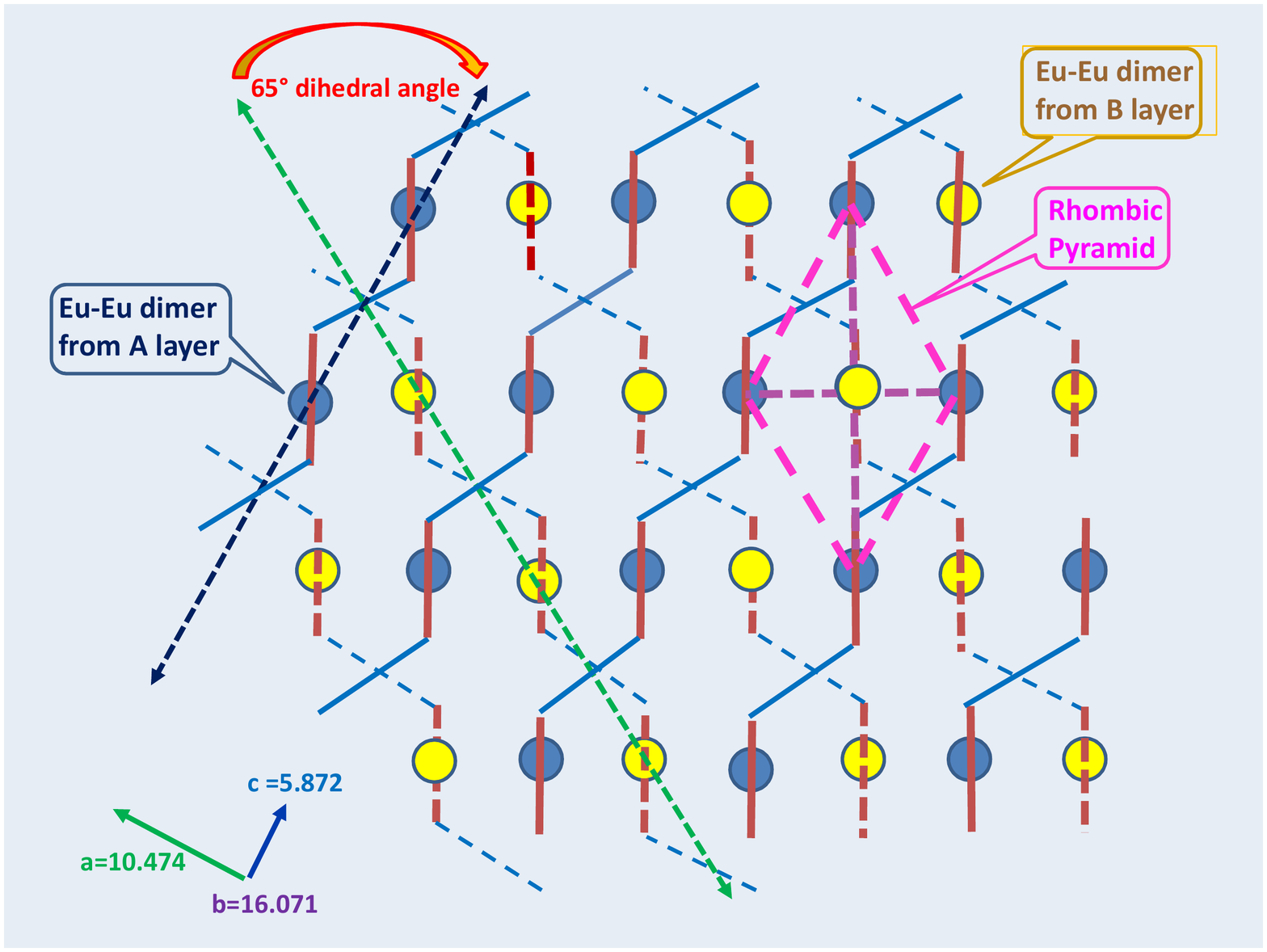}
\caption{Reformulation of Eu$_2$Pd$_2$Sn magnetic structure based on dimer's zig-zag chains included in two consecutive layers, represented by blue circles for layer A and yellow ones for layer B. Long dashed arrows (blue and green) indicate the 65° dihedral angle between chains \cite{ours}. The rhombic piramid (magenta-dashed lines) indicate the triangular configuration between dimers belonging to two consecutive layers. \label{F11}}
\end{center}
\end{figure}

As shown in Fig.~\ref{F1} \cite{ours, Doverbratt} and discussed in the introduction, the crystalline structure of Eu$_2$Pd$_2$Sn is particularly complex, making the study of the involved magnetic interactions quite difficult. 
Furthermore, once established that Eu-chains transform into dimer-chains, the topology of this compound is 
modified by the presence of these condensed quasiparticles. With the aim to review the consequences of this new landscape, we reformulate the Eu$_2$Pd$_2$Sn interactions pattern in Fig.~\ref{F11} highlighting the dimer  
chains in the puckered hexagonal network of Eu atoms. Notice that, while dimer chains are depicted along the 'c' axis in layer A, the upper ones (belonging to layer B) are rotated in 65°. 

According to these considerations and taking into account the observed magnetic behavior above $T_N$, it is evident that the strongest (FM) interaction occurs between Eu-NN: $K_x > 0$, within the Eu zig-zag chains along the 'c' crystalline direction. This interaction is responsible 
for the Eu-Eu dimers formation becuase of the strong Eu-Eu electronic overlap. The second magnetic interaction: $K_y < 0$, becomes relevant at the AF ordering temperature because it connects neighboring Eu-dimers along the chain and drives the formation of 
modulated or helical ordering. 
Based on the dimers zig-zag chains extended along the ‘c’ direction (see Fig.~\ref{F1}c) and the mentioned similarities with thermodynamic and magnetic studies on EuNiGe$_3$, an incommensurate helicoidal ordering appears to be 
congruent with a possible magnetic structure for Eu$_2$Pd$_2$Sn at zero field.

As it was described in  Fig.~\ref{F1}a, successive layers are stacked along the ‘b’ crystalline direction in an ABA’B’ … sequence, 
with the Eu dimers chains disposed in a dihedral angle of ca. 65° between neighboring AB layers once projected on the ‘ac’ plane, see the dashed dark-blue and green arrows in  Fig.~\ref{F11}. In that figure also the B layer 
projection is included, with Eu dimers represented by blue circles for layer A and yellow for layer B.  As it can be seen, neighboring dimers belonging to the same plane, form a regular triangular 
lattice and build up a rhombic pyramid with the first neighbor of the upper plane in its vertex. The pyramid itself displays four triangular faces that, together with the triangular network in plane, provide 
geometrical conditions for magnetic frustration. The equivalent rhombic pyramid is formed between 
A’ and B’ layers, but shifted in the ‘ac’ plane as depicted in Fig.1-II of Ref \cite{ours} .

\subsection{Possible formation of a Skyrmions phase}

It is well known that, under cooling, frustrated systems are compelled to search for alternative ground states with lower degeneracy in order to reduce its highly entropic GS. That constraint favors to access to exotic 
phases which are not reached by standard magnets that can order at higher temperatures  \cite{Constraints}. Among a variety of possible magnetic configurations, skyrmion vortex lattices provide an 
alternative to build up 
some type of coherent structure occurring in frustrated systems. Skyrmions are described as rather ubiquitous topological magnetic structures observed in several magnetic materials without inversion symmetry \cite{Fert}.

These topologically stable structures appear as triangular crystals of vortex lines parallel to the field direction, which are 
manifested as small pockets in the magnetic phase diagram of non-centrosymmetric magnets \cite{Xtian}.
Eu$_2$Pd$_2$Sn fulfils a number of conditions for the formation of those skyrmion structures like: magnetic frustration,  modulate type of propagation vector along one direction which differs around 120° (as the 
complementary angle to the observed ca.65°) with the neighboring chain, and it shows one of those eventual pockets between $B_{C1}$ and $B_{C2}$ associated to a critical point CP$_1$ in its phase diagram (see Fig.~\ref{F10}). In Ref. \cite{Xtian} a series of alternative magnetic phase diagrams is presented, 
obtained by Montecarlo simulations and 
computed according to selected values of neighboring magnetic couplings which produce different propagation vectors. Among the proposed phase diagrams, the one with ''moderate adjacent interlayer exchange and zero interaction between NNN layers'' (Fig.1a in Ref. \cite{Xtian}) also presents a tetracritical point among the similarities with that obtained for Eu$_2$Pd$_2$Sn.

\subsection{Tentative recognition of the magnetic phases}

The reformulation of the magnetic configuration based on the presence of Eu dimers presented in Fig.~\ref{F11}, the detailed comparison with EuNiGe$_3$ single crystals behavior and the possibility 
of a skyrmion phase formation, allow to proceed towards a reliable identification of most of 
the magnetic phases presented in Fig.~\ref{F10}.

At zero and low field, the quasi-paramagnetic phase above the Neel temperature is marked by the dimers formation. Below the transition an incommensurate AF-ICM phase appears, which transforms into an helicoidal 
incommensurate AF-ICH type of order. Around $B\approx 1$\,T two critical points are found. 
One, CP$_1$ at $T\approx 7$\,K, with characteristics of tetracriticality, marks the upper temperature limit of the proposed skyrmions lattice phase. The other,  CP$_2$ at $\approx 12$\,K, marks the limit of the $T_N(T,B)$ transition which exhibits signs of first order. 
Between B$_{C2}$ and B$_{C3}$ critical fields there is an unidentified ('unknown') phase due to the lack of enough information about its nature. Above B$_{C3}$ the magnetic phases which be more likely  identified as: partially 
FM-PP and fully FM-FP polarized ferromagnets. It is evident that microscopic investigation, like neutron diffraction or Moessbauer spectroscopy, applied on single crystal samples are required to confirm these magnetic structures.

\section{Conclusions}

The Eu$^{2+}$ network in Eu$_2$Pd$_2$Sn can be represented as formed by zig-zag chains located in puckered elongated hexagons. 
A peculiar property of this compound is the reduced atomic spacing between Eu atoms into the zig-zag chains in comparison with that evaluated for pure Eu$^{2+}$. As a consequence a strong overlap with the NN electronic cloud is expected, to which the formation of Eu-dimers quasiparticles starting to condense at quite high temperature ($\approx 70$\,K) can be attributed. 

Such dimers formation, with a very large $J_D=7$ quantum number, allows to understand the deviation of the magnetic susceptibility from the high temperature C-W law, the magnetization deviation respect to  
the $B_{7/2}$ Brillouin function at high temperature, and the entropy accumulated at the ordering temperature $T_N$. 
The presence of these quasiparticles requires a reformulation of the magnetic structure of the compound, with the expected consequences in the role of the magnetic interactions like the transformation from a FM dominant interaction at high temperature to an AF one around $T_N$.

A rich magnetic phase diagram is obtained from the analysis of the temperature derivatives of the magnetic parameters and the field dependence of the specific heat. From this study, two critical points are recognized. The 
lack of microscopic knowledge of the magnetic 
nature of those phases can be eluded by comparing its behavior with that of a mirror compound EuNiGe$_3$ investigated on single crystals samples. From that comparison, below the AF transition at 13.3\,K Eu-dimers form 
incommensurated AF chains whose topological characteristics provide the conditions for geometric frustration between neighboring dimers of a layer and those of the subsequent one within a rhombic pyramid configuration 

The possible formation of a skyrmion lattice arises from the presence of those magnetically frustrated pockets, according to alternative phase diagrams proposed by theoretic studies on hexagonal structures which present similar interactions pattern. Such possibility requires to be confirmed by microscopic magnetic studies performed on single crystal samples.   

\section*{Acknowledgements}
This research work is part of
the Project implementation: University Science Park TECHNICOM for Innovation
Applications Supported by Knowledge Technology, ITMS: 313011D232, supported by
the Research \& Development Operational Programme funded by the ERDF; and
also by VEGA1/0705/20, 1/0404/21.


\begin{thebibliography}{00}
\bibitem{jump} From mean field theory, the specific heat jump at $T_N$ is given by \cite{MF}:
$\Delta C(T_N)$=5/2R$[(2J+1)^2 -1]/[(2J+1)^2 +1]$\\
Notice that in the figure the specific heat is presented in [J/mol K] units whereas for this formula $\Delta C(T_N)$ is expressed in [J/atom K] units.
\bibitem{bkgr} The full $C_{bk}(T) $curve was obtained by fitting $C_{4f}$ within the $5 \leq T \leq 22$\,K range using as fitting function: $C_{bk}=193/[(T-10.8)^2+14]+3*atan((T-11.5)/7)+8$.
\end{thebibliography}
\end{document}